\documentclass[aps,pra,twocolumn,superscriptaddress]{revtex4-1}
\usepackage[utf8]{inputenc}
\usepackage{braket,textcomp}
\usepackage{color}
\usepackage[dvipsnames]{xcolor}
\usepackage[breaklinks, colorlinks=true, urlcolor=blue, anchorcolor=blue, citecolor=blue, filecolor=blue, linkcolor=blue, menucolor=blue, linktocpage=true, pdfproducer=medialab, pdfa=true]{hyperref}
\usepackage{amsbsy,amssymb,amsmath,bbm,mathcomp}
\usepackage{times}
\usepackage{float}
\usepackage{graphicx}
\newcommand{{\sv}}{${\cal S}$}
\newcommand{{\A}}{A}
\newcommand{{\B}}{B}

\newcommand{\eli}[1]{{\color{Emerald} {\bf eliana: }{#1}}\\}
\newcommand{\alex}[1]{{\color{Violet} {\bf alessandro: }{#1}}\\}

\newcommand\lsim{\mathrel{\substack{\textstyle<\\[-0.3ex]\textstyle\sim}}}
\newcommand\gsim{\mathrel{\substack{\textstyle>\\[-0.3ex]\textstyle\sim}}}
\newcommand{\beq}{\begin{equation}}
\newcommand{\eeq}{\end{equation}}
\begin{document}
\title{Critical slowing down and entanglement protection}
\author{Eliana Fiorelli}
\address{Dipartimento di Fisica e Astronomia, Universit\`a di Firenze, I-50019, Sesto Fiorentino (FI), Italy}
\address{INFN, Sezione di Firenze, I-50019, Sesto Fiorentino (FI), Italy}
\address{School of Physics and Astronomy, University of Nottingham, Nottingham, NG7 2RD, UK}
\address{Centre for the Mathematics and Theoretical Physics of Quantum Non-equilibrium Systems, University of Nottingham, Nottingham NG7 2RD, UK}
\author{Alessandro Cuccoli}
\address{Dipartimento di Fisica e Astronomia, Universit\`a di Firenze, I-50019, Sesto Fiorentino (FI), Italy}
\address{INFN, Sezione di Firenze, I-50019, Sesto Fiorentino (FI), Italy}
\author{Paola Verrucchi}
\address{ISC-CNR, UOS Dipartimento di Fisica, Universit\`a di Firenze, I-50019, Sesto Fiorentino (FI), Italy}
\address{Dipartimento di Fisica e Astronomia, Universit\`a di Firenze, I-50019, Sesto Fiorentino (FI), Italy}
\address{INFN, Sezione di Firenze, I-50019, Sesto Fiorentino (FI), Italy}
\date{\today}
\begin{abstract}
We consider a quantum device $D$ interacting with a 
quantum many-body environment $R$ which features a second-order phase 
transition at $T=0$. Exploiting the description of the critical slowing 
down undergone by $R$ according to the Kibble-Zurek mechanism, we 
explore the possibility to freeze the environment in a configuration 
such that its impact on the device is significantly reduced. Within this 
framework, we focus upon the magnetic-domain formation typical of the 
critical behaviour in spin models, and propose a strategy that allows 
one to protect the entanglement between different components of $D$ from 
the detrimental effects of the environment.

\end{abstract}
\maketitle
%%%
\section{INTRODUCTION}

In the last decades, studies on how to manipulate quantum systems have 
boosted the scientific community's confidence in regard to the possible 
realization of quantum devices. These are usable apparatuses whose 
operating principles are based on genuinely quantum properties, amongst 
which entanglement between components is key to outperforming classical 
machines. Given that an apparatus is usable if some external control can 
be exerted on the state and evolution of its components, the description 
of a quantum device must envisage the existence of at least another 
system, that enforces such control by interacting with the device 
itself. This means that a quantum device is open to the external world 
by definition, and it is not a stretch to name "environment" whatever 
influences its behaviour from outside. For this reason, the analysis of 
how quantum devices work implies the study of how open quantum systems 
evolve\cite{BreuerP02,RivasH12,LoFranco15,Koch16,DuffusDE17,BullockEtal18, GiorgiEtal13,Vicari18,BellomoEtal10}.

In fact, it is one of the most challenging tasks of quantum technologies 
that of allowing quantum devices to be "open" and yet to properly 
function \cite{LoFrancoEtal14}: quantum properties are fragile and vulnerable to the 
environmental impact, and strategies for their protection most often 
imply either the suppression of the interaction between environment and 
device (which is never exactly achievable) or a very detailed design of 
their couplings (which is usually an experimentally arduous task).

In this work we aim at understanding if a quantum device $D$ can 
be protected by intervening on some properties of its environment 
$R$, without neither quenching the 
interaction between $D$ and $R$ nor giving it too peculiar a form.
To this aim, we specifically consider the case when $R$ is a quantum
many-body system featuring a second-order phase transition, 
and investigate the possible consequences of a critical 
behaviour of $R$ on the entanglement between different 
components of $D$. Reason for this choice is the possible exploitation
of the critical slowing down leading to the 
Kibble-Zurek mechanism (KZM) in order to freeze the environment in a 
configuration that is as harmless as possible for the device. 

In order to focus this argument, we first notice that any entanglement 
between $D$ and $R$ (hereafter dubbed "external") is useless as far as 
the device functioning is concerned, and its buildup inevitably goes  
with damages of that between different components of $D$ (hereafter 
dubbed "internal"), which is the useful one. A naive strategy for 
protecting the latter by minimizing the former is to deal with an 
environment that behaves almost classically \cite{FiorelliCV19}, which is to say it can only be 
weakly entangled with any other system. Referring to the case of a 
magnetic environment, which is what we will hereafter do, an 
almost-classical $R$ can be obtained in the form of a system with a 
large value of the spin ${\cal S}$\cite{Lieb73}; however, the effect
of one such system upon each component of $D$ could be so 
prevailing to squash the fragile 
quantum machineries that make the device function, up to the point of 
making its state always separable, as if its components were not part of 
a unique, composite, system $D$.

We therefore propose another strategy, based on the
magnetic-domain formation which is typical of the critical 
behaviour of many spin-models. In fact, each domain is a 
large-${\cal S}$ system and yet different domains do not point into the same 
direction, which should result into an overall weaker
effect of $R$ on the device components. Moreover, the dynamics 
of magnetic domains can be significantly slowed down in the vicinity of
a second-order phase transition, which might also help protecting the 
internal entanglement.

The paper is organized as follows: In Sec.\,\ref{s.model} we define 
the model of the quantum device and its environment, with Secs.\,\ref{ss.critical} 
and \ref{ss.KZM}  devoted to a brief descritpion of the critical slowing down 
and the Kibble-Zurek mechanism, respectively. 
In Sec.\,\ref{s.essential_toolbox} we introduce the tools used
in Secs.\,\ref{s.para} and \ref{s.diabatic} to study the evolution of 
the overall model. The dynamics of the entanglement between the device 
components is described in Sec.\, \ref{s.entanglement}. Our 
results are presented and discussed in Sec.\,\ref{s.Results}, and
conclusions are drawn in the last section.

\section{Model}
\label{s.model}

\begin{figure}[H]
\centering
\includegraphics[width=0.6\linewidth]{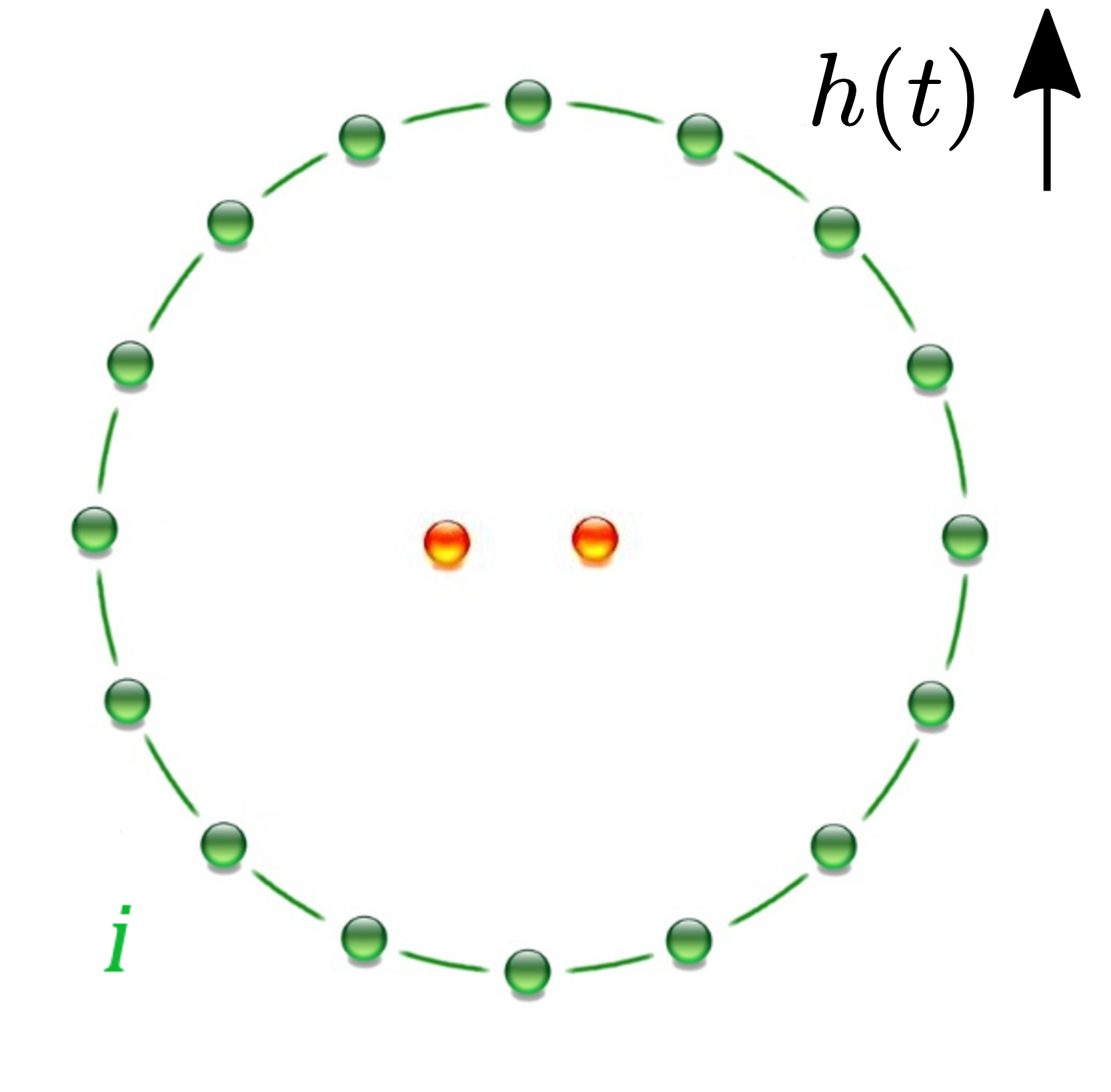}
\caption{\small
{Schematic representation of the ring of spins with a central qubit pair.}}
\label{f.spin_ring}
\end{figure}

We consider a "device-plus-environment" quantum system, 
$\Psi=D+R$, 
where the device is a qubit pair, $D = {\A} + {\B}$, and 
the environment is a ring $R$, made of $N$ spin-$\frac{1}{2}$ particles, 
as shown in Fig.~\ref{f.spin_ring}.
Each qubit is described by the Pauli operators
$\hat{\boldsymbol{\sigma}}_*$, with 
$\left[\hat{\sigma}_*^{\alpha},\hat{\sigma}_*^{\beta} \right] = 
i2\epsilon_{\alpha\beta\gamma}\hat{\sigma}_*^\gamma$, 
$\alpha(\beta,\gamma)=x,y,z$, and $*={\A},{\B}$, while
elements of $R$ are described by operators 
$\hat{\mathbf s}_i$, with 
$\left[\hat s_i^{\alpha},\hat s_j^{\beta} \right] = 
i\epsilon_{\alpha\beta\gamma}\delta_{ij}\hat s_i^{\gamma}$,
$|\hat{\mathbf s}_i|^2=\frac{3}{4}$,
$i(j)=1,...N$, and 
periodic boundary conditions enforced,
$\hat{\mathbf s}_{N+1}=\hat{\mathbf s}_1$.

As we want $R$ to feature a second-order phase transition that survives 
the lowering of temperature (so that we can reduce the thermal effects 
without modifying our setting), we focus upon Quantum Phase 
Transitions (QPT), which are second-order phase transitions occurring at 
zero temperature, under the tuning of some model parameter. To this 
respect notice, however, that quantum critical properties survive at 
sufficiently low and yet finite temperatures, which makes the following 
analysis amenable to experimental investigation.
The $N\to\infty$ limit underlying the occurrence of any genuine 
phase-transition is implemented by combining a large value of $N$ with 
the periodic boundary conditions inherent in the ring geometry. 

In the above general framework we specifically choose $R$ to be 
described by a prototypical spin model for one-dimensional 
QPT. 
As for the two qubits, they are 
coupled with 
each component of the ring via a ZX ferromagnetic exchange but
do not interact amongst themselves and they are
not subject to the transverse field that drives the QPT.
We will comment upon these choices in the concluding section.

The dimensionless Hamiltonian of the system is 
\begin{equation}\label{e.Htot} 
\hat{H}=\hat{H}_{\rm 
I}-\frac{g}{2}(\hat\sigma_{\A}^z+\hat\sigma_{\B}^z)
\sum_{i=1}^{N}(\hat{s}_{i}^{+}+\hat{s}_{i}^{-}),
\end{equation} 
with
\begin{equation}\label{e.Hring} 
\hat{H}_{\rm I}=
-\sum_{i=1}^{N}\hat{s}_{i}^{x}\hat{s}_{i+1}^{x} 
-h(t)\sum_{i=1}^{N}\hat{s}_{i}^{z} 
\end{equation}
the Hamiltonian of the ring, where we have chosen a ferromagnetic 
Ising interaction, whose exchange integral $J$ sets the energy scale (i.e. 
the physical Hamiltonian of the model is the above
dimensionless one times the actual exchange integral 
$J$).
The coupling $g$ is 
positive, and 
$h(t)$ accounts for the presence of a time-dependent transverse (i.e. 
pointing in the positive $z$-direction) magnetic field that drives the 
QPT. We will also consider the case of constant field.

As for the initial state of the system, we will take it separable as far 
as the partition $D+R$ is concerned,
\beq
\ket{\Psi(0)}=\ket{D}\otimes\ket{R}~,
\label{e.ini_state-separable}
\eeq
where the state of the qubit pair is
\beq
\ket{D}=\sum_\gamma c_\gamma\ket{\gamma}~,
\label{e.ini_state-qubits}
\eeq
with
$\{\ket{\gamma}\}_{{\cal H}_D}$ the four eigenstates of 
$(\hat{\sigma}^z_{\A}+\hat{\sigma}^z_B)$, generically labelled by the 
index $\gamma=1,...4$; the coefficients 
$c_\gamma$ are complex numbers 
and are different from zero for at least two different $\gamma$s, in 
order to ensure that $A$ and $B$ are entangled.

\subsection{Critical behaviour of the ring}
\label{ss.critical}
The Hamiltonian $\hat{H}_{\rm I}$ 
defines the one-dimensional quantum Ising model in a transverse 
field (QIf), which is a paradigmatic example of a system undergoing 
a QPT~\cite{Sachdev11}. The transition occurs due to the competition 
between the 
action of the 
external field, that supports independent alignment of each spin along 
the $z$ direction,
and 
the exchange coupling amongst adjacent spins, that favours their 
being parallel to each other and all pointing in the $x$ 
direction.  The control parameter driving the transition is 
the field $h$, with the QPT located at $h=1$;
the region where critical behaviours are observed is 
usually dubbed {\it critical region}.
For the sake of clarity, in this section we will use the parameter
\begin{equation}
\epsilon:=h-1,
\label{e.epsilon}
\end{equation}
and set the QPT at $\epsilon_c=0$.

The order parameter for the QIf is the $x$ component of the 
magnetization
\begin{equation}
\frac{1}{N}\sum_j\braket{\hat{s}_{j}^{x}}=\braket{\hat{s}_{i}^{x}}\equiv 
m~,~\forall i~,
\label{e.magnetization}
\end{equation}
where by $\braket{~\cdot~}$ we indicate the expectation value upon 
the 
translationally invariant ground state;
$m$ is finite in the ordered phase ($\epsilon<0$), and it vanishes in
the disordered one ($\epsilon>0$). 
In the critical region, the related correlation functions
$\chi_r:=\braket{\hat{s}_{i}^{x}\hat{s}_{i+r}^{x}}$ behave according to
\begin{equation}
\chi_r-m^2\sim e^{-\frac{r}{\xi}}~,
\label{e.correlations}
\end{equation}
where $\xi$ is the correlation length, that diverges at the transition 
as
\begin{equation}
\xi \sim \frac{\xi_{0}}{|\epsilon|^{\nu}}~, 
\label{e.csi}
\end{equation} 
with $\nu>0$ the corresponding critical exponent, and $\xi_{0}$ a 
non-universal length scale. 
We notice that $\xi$ is sometimes dubbed "healing" length, 
to indicate that it sets the scale upon which
$\braket{\hat s^x_i}$ heals in space, returning to the
spatially homogeneous value $m$ after having been affected by 
a local fluctuation. 
A similar concept can be introduced for describing the way the system 
reacts to instantaneous, i.e. local in time, fluctuations. 
This leads to the introduction of a quantity called
relaxation, or "reaction" time $\tau$, that sets 
the time-scale upon which the relevant quantities 
settle, after the control parameter has varied instantaneously.
The reaction time is also known to diverge at the transition, according 
to
\begin{equation}
\tau\sim\frac{\tau_{0}}{|\epsilon|^{\nu z}}~,
\label{e.tau}
\end{equation} 
with $z>0$ the so called "dynamical" critical exponent, and $\tau_0$ a 
non-universal time-scale.
It is worth mentioning that the product $\nu z$ also
rules the critical vanishing of the gap $\Delta$ between the 
ground state energy and that of the first-excited one, 
$\Delta\sim|\epsilon|^{\nu z}$, signalling the most relevant relation 
between such vanishing and the occurrence of the QPT itself.
Without further commenting upon this point, which is extensively 
discussed in the literature, let us specifically address
Eq.\eqref{e.tau}.

A divergent relaxation time implies an 
extremely slow dynamics of the system as a whole, with fast 
fluctuations occurring only locally without any significant effect on 
the global scale set by the correlation length. 
This phenomenon, which is usually referred to as "critical slowing 
down", is evidently intertwined with the divergence of the correlation 
length, if only for the fact that both Eqs.~\eqref{e.csi} and 
\eqref{e.tau} describe a power-law divergence at 
$\epsilon=\epsilon_c=0$.
On the other hand, a finite relaxation time is key to the definition 
of adiabaticity, i.e. the distinctive feature of dynamical regimes where 
the system changes its state (or configuration, in the classical 
case), after the variation of relevant parameters, in a 
time-interval that is much shorter than the time-scale of the
variation itself.
Therefore, we expect that the divergence of $\tau$ at $\epsilon=0$,
and the consequent critical slowing down,
be related to the onset of a non-adiabatic regime, sometimes called 
"diabatic", which is indeed at the hearth of the Kibble-Zurek 
mechanism (KZM) described below.

\subsection{The Kibble-Zurek mechanism}
\label{ss.KZM}

An exact analytical description of the dynamical evolution of a 
many-body quantum system which is driven across its phase 
transition is an unattainable task, due to the very same
reason why the transition occur, i.e. the presence of 
terms in the system Hamiltonian that do not commute, not even at 
different times. From a numerical viewpoint, the situation is 
equivalently intractable, even in a classical system, because of the several 
divergences that characterize whatever critical behaviour. However, in 
the same spirit that allows one to derive and use equations such as 
Eqs.~\eqref{e.csi} and \eqref{e.tau}, 
it is possible to elaborate on criticality to get an effective 
description of the process through which a phase transition dynamically 
happens. This is how Kibble and Zurek built up a paradigm for 
describing out-of-equilibrium dynamics around a continuous phase 
transitions, today known as the "Kibble-Zurek mechanism" (KZM).
The theory, was initially proposed by Kibble \cite{Kibble76} within 
the cosmological context, later extended by Zurek \cite{Zurek85,Zurek96} 
to condensed matter systems, and finally to QPT \cite{ZurekDZ05, 
Dziarmaga05, Dziarmaga10,delCampoZ14, SilviEtal16, FubiniFO07}.
The mechanism takes different forms depending on the model-Hamiltonian 
and the functional time-dependence of the control parameter.

In this subsection we describe the KZM for the QIf when the transverse 
field varies linearly in time, 
\begin{equation}
h(t)=h_0-vt~,
\label{e.h(t)}
\end{equation}
with positive velocity $v$;
$1/v$ is referred to as the "quench time", suggesting that
the transition is crossed by lowering the field, i.e. moving
from the disordered to the ordered phase.
In fact, this is the process to which we will refer in this work, with 
$h_0>1$ to set the model 
in the disordered phase when the process starts.

The control parameter in Eq.~\eqref{e.epsilon} is
\begin{equation}
\epsilon(t)=h(t)-1=(h_0-1)-vt~,
\label{e.epsilon(t)}
\end{equation}
that embodies the definition of a critical time
\begin{equation}
t_c=\frac{h_0-1}{v}
\label{e.t_c}
\end{equation}
after which the QPT is reached; more generally, the 
time left before the transition is crossed is 
\begin{equation}
\delta(t)=t_c-t=\frac{h_0-1}{v}-t~.
\label{e.delta(t)}
\end{equation}
The key observation leading to the KZM is that, due to the 
divergence of the reaction time Eq.~\eqref{e.tau},
there certainly exists a finite time-interval where
\begin{equation}
\delta(t)\le\tau~,
\label{e.dia_cond}
\end{equation}
meaning that before the system has reacted globally to the 
control-parameter variation, the critical point has already been 
reached, a situation which is evidently inconsistent with whatever 
adiabatic-like dynamics. 
In fact, if Eq.~\eqref{e.dia_cond} holds, 
the system cannot meekly adapt to the variation of the control 
parameter but rather gets stuck on a configuration that is qualitatively 
the one taken when $\delta(t)=\tau$, i.e. at the time 
\begin{equation}
\bar{t}=\frac{h_0-1}{v}-\sqrt{\frac{1}{2v}}~,
\label{e.bart}
\end{equation}
where we have used Eqs.~\eqref{e.tau} and 
\eqref{e.epsilon(t)}
with $\nu=z=1$ and $\tau_0=\frac{1}{2}$, which are the due values for the 
QIf.

From the above description the "diabatic" dynamical regime is set in the 
time-interval
\begin{equation}
\bar t\le t\le 2t_c-\bar t~.
\label{e.diabatic}
\end{equation}

\begin{figure}[H]
\centering
\includegraphics[width=\linewidth]{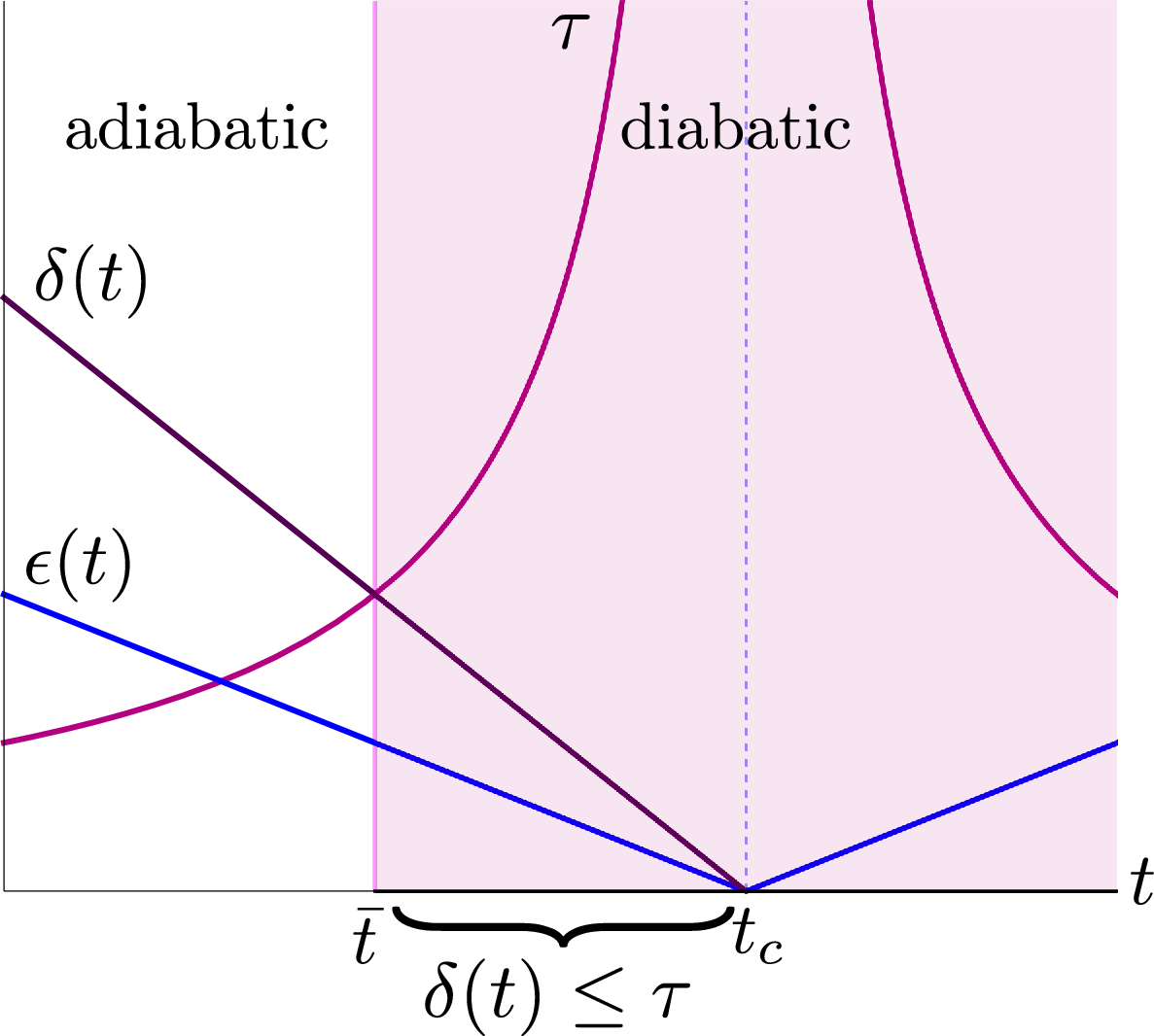}
\caption{Schematic representation of the KZM. See text below Eq. (\ref{e.diabatic})
for a detailed description.  
}
\label{f.kzm}
\end{figure}

The aforesaid process can be graphically depicted as in Fig. 
(\ref{f.kzm}). The magenta line represents the reaction time $\tau$
that diverges at $t_c$, when $\epsilon(t_c)=0$ and the critical 
point is reached. 
The purple 
line is $\delta(t)$ as from Eq.~(\ref{e.delta(t)}), while the
blu one is $\epsilon(t)$, 
Eq.\,(\ref{e.epsilon(t)}).
The shaded area is where $\delta (t) \leq \tau$.

In fact, the KZM goes beyond the above phenomenology, and describes its 
implications as far as the dynamics of the system is concerned. In the 
remaining part of this section, we discuss these implications for
the QIf in the disordered phase, aimed at devising 
approximations to be used in the diabatic regime.

Referring to the process as represented in Fig.~\ref{f.kzm}, 
we know that for large fields, despite the ring being in its disordered 
phase as far as the spin correlations along the $x$ direction are 
concerned, its ground state is "ordered", with all the spins aligned 
along the $z$ direction, though independent from each other. 
Consistently with the usual terminology, we will understand that in the 
disordered non-critical phase, the ring exhibits a "paramagnetic" 
behaviour.

Once the quench starts
the dynamics is still adiabatic, with the magnetization along the
$z$ direction that decreases with time,
as far as the reaction time is smaller 
than $\delta(t)$ (i.e. for $t\lsim\bar t$).
However, blocks of dynamically correlated spins, hereafter dubbed "domains",
begin to appear.
If the exponential behaviour \eqref{e.correlations} 
has already set in, a correlation length exists and it makes 
sense to take the length of the above domains just of the
same order of magnitude.

When the QPT gets closer, and the reaction time becomes much longer than 
$\delta(t)$ (i.e.  for $t\gsim\bar t$), adiabaticity is lost:
The ring has no time to conform its state to
the instantaneous ground-state of the time-dependent Hamiltonian, 
and it gets stuck into the state where it was at $t=\bar t$, 
with domains of average length 
$\xi[(\epsilon(\bar t)]:=\xi_{d}$. 
Due to the 
homogeneity of 
the Ising coupling along the ring, 
these domains require a time which is proportional to $\xi_d$  
to estabilish dynamical correlations amongst themselves.
On the other hand, at $t=\bar t$ the system is in its critical region, 
meaning that $\xi_{d}$ is very large. Therefore,
different domains cannot 
be causally connected and can be effectively described as non-interacting.

This is a relevant point in Sec. \ref{s.diabatic}, where the formation 
of effectively non-interacting domains allows us to describe $R$ in 
terms of large, independent spins.

\subsection{Weak-coupling constraint}
\label{ss.weak-coupling}

The phenomenology described in the subsections \ref{ss.critical} 
and \ref{ss.KZM} refers to the ring as if it was not 
interacting with the two qubits, i.e. as if $g=0$ in the 
Hamiltonian \eqref{e.Htot}. On the other hand, we aim at exploiting the 
KZM to control the dynamics of the complete model, with $g>0$. 

This is made possible by 
enforcing a weak-coupling constraint
\begin{equation}
g\ll 1 ~~~{\rm and} ~~~ g\ll h(t)
\label{e.weak-coupling}
\end{equation}
throughout the rest of this work. 
This condition has no implications on the description of the 
critical behaviour, which occurs when $|h(t)|\sim 1$, but it definitely 
rules out the 
region where the ring becomes effectively ordered due to the vanishing 
of $h(t)$. Therefore, to avoid inconsistencies w.r.t. this point, our 
analysis will exclusively concern the disordered 
phase $h(t)\ge 1$, where conditions 
\eqref{e.weak-coupling} can be safely assumed.  

This will be used in Sec. \ref{s.para} and \ref{s.diabatic}, in order to 
get an effective propagator and hence the evolved 
state of the system, both in the paramagnetic and the diabatic setting.

\section{Strategy and essential toolbox}
\label{s.essential_toolbox}

In this section we explain our goal, and provide the reader with the 
essential tools we have used to accomplish it.

Referring to the possible strategies to protect internal entanglement 
mentioned in the Introduction, we will compare the way the 
entanglement 
between $A$ and $B$ 
decreases after the interaction with $R$ is switched on, in two 
different settings, both relative to the disordered, $\epsilon > 0$ phase.

Firstly we will consider the dynamics of the model for a constant 
large value of 
$h$, so as to set the ring far from its
critical region; in this case we expect it 
to behave as an almost classical paramagnet, acting upon $D$ as if 
it were one 
single system with a very large spin ${\cal S}$, pointing in the 
direction of the field. The overall evolution of the system will be 
effectively ruled by the coupling between $D$ and $R$ only, and we will 
refer to this setting as the "paramagnetic" case.

Secondly we will set the ring well within the critical 
region and drive 
it 
into the diabatic regime by quenching the magnetic field as
$h(t)=h_0-vt$, with $h_0\gsim 1$. The time-dependence of the 
field will enter 
the evolution of the system (with the KZM playing an essential role 
in effectively describing it), and  we will refer 
to this setting as the "diabatic" case.

In both settings we will study how the initial state 
\eqref{e.ini_state-separable} changes 
under the action of the propagator defined by the Hamiltonian 
\eqref{e.Htot}; this will allow us to obtain the evolved state of $D$ (a 
mixed state due to the generation of external entanglement) 
via the partial trace over the Hilbert space of $R$; the 
internal entanglement dynamics will be finally analysed in 
terms of the time dependence of the concurrence \cite{Wootters98} 
between the two qubits.

Despite the peculiar features of the paramagnetic and 
diabatic regimes, the coupling between $R$ and $D$ makes 
it impossible to exactly determine the evolution of the state 
\eqref{e.ini_state-separable}.
This is due to the commutation rules obeyed by 
spin operators, that most often prevent one from getting closed 
factorized expressions for the propagators by the Zassenhaus 
formula\cite{Zassenhaus39}, 
i.e. the dual of the Baker-Campbell-Hausdorff one.
Moreover, we need to give the initial state of the 
ring, $\ket{R}$, an explicit form, which 
is a non trivial problem per s\'e, tantamount to 
determine the ground state of an interacting, possibly 
critical, many-body system.

As a matter of fact, in Secs.\ref{s.para} and 
\ref{s.diabatic} we will factorize the propagator 
$\exp\{-it\hat{H}\}$ by the Zassenhaus 
formula for spin operators, possibly with large-${\cal S}$ , and apply 
it 
to the initial state \eqref{e.ini_state-separable}, with $\ket{R}$  
described by spin-\sv$~$Coherent States (\sv CS).  Therefore, 
in the 
following 
subsections we introduce the Zassenhaus formula, explain how the 
large-${\cal S}$ condition is formally implemented, and briefly recall 
essential facts about \sv CS.
 
\subsection{Zassenhaus formula}
\label{ss.zassenhaus}
Given two non-commuting operators $\hat{X}$ and $\hat{Y}$, the 
Zassenhaus formula reads
\begin{equation}\label{Zassenhaus}
e^{\lambda(\hat{X}+\hat{Y})}
=e^{\lambda \hat{X}}e^{\lambda \hat{Y}} 
e^{\lambda^{2}C_{2}(\hat{X},\hat{Y})}\cdots 
e^{\lambda^{n}C_{n}(\hat{X},\hat{Y})} \cdots,
\end{equation}
where the operators $C_{n}(\hat{X},\hat{Y})$ have been recently 
expressed\cite{CasasM12} as 
\begin{align}
&C_{n+1}(\hat{X},\hat{Y})= \frac{1}{n+1}\times\\
&\times\!\!\!\!\!\!\!\underset{{(i_{0},...,i_{n}) \in 
\mathcal{I}_{n}}}{\sum}\left[
\prod_{k=0}^{n}\frac{(-1)^{i_k}}{i_k!}\right]
{ad}_{C_{n}}^{i_{n}} 
\cdots {ad}_{C_{2}}^{i_{2}}{ad}_{\hat{Y}}^{i_{1}}
{ad}_{\hat{X}}^{i_{0}}\hat{Y}~,
\label{e.Cn+1}
\end{align}
where
\begin{equation}
{ad}_{\hat{X}}^{0}\hat{Y}=\hat{Y},\qquad {ad}_{\hat{X}}^{k}\hat{Y}=  
[\underbrace{\hat{X},[\hat{X},...[\hat{X}}_{k\,\mathrm{times}},\hat{Y}]...]]~,
\label{e.adjointaction}
\end{equation}
with $\mathcal{I}_{n}$ the set of $(n+1)$-tuples of non-negative 
integers $(i_{0},i_{1},...,i_{n})$ satisfying the conditions:
\begin{equation}
 i_{0}+i_{1}+2i_{2}+...+ni_{n}=n, \quad k+1\leq i_{0}+...+ki_{k}\quad 
\forall k \leq n-1~.
\label{e.cond-nupla}
\end{equation}
Equivalently, the "left-oriented" version of (\ref{Zassenhaus}) reads
\begin{equation}\label{e.Zassenhaus_left}
e^{\lambda(\hat{X}+\hat{Y})}=\cdots e^{\lambda^{n}\tilde{C}_{n}(\hat{X},\hat{Y})}\cdots e^{\lambda^{2}\tilde{C}_{2}(\hat{X},\hat{Y})}e^{\lambda \hat{Y}}e^{\lambda \hat{X}},
\end{equation}
with $\tilde{C}_{n}=(-1)^{n+1}C_{n}$, $n \geq 2$.

Eq.~\eqref{e.adjointaction} makes it clear that whenever 
$[\hat X,\hat Y]$ is not a number, $\exp\{\lambda({\hat X}+{\hat Y})\}$ 
factorizes into a product of infinite terms that contain increasingly 
nested commutators.
However, when $\hat{X}$ and $\hat{Y}$ are spin operators describing 
a system with a large value of ${\cal S}$, we can obtain a reasonable 
approximation by the following argument.
\subsection{Large $\cal S$}
\label{ss.largeS}
When dealing with Hamiltonians that contain terms $g\hat S^\alpha$ 
with $g$ 
some coupling constant, as in Eq.~\eqref{e.Htot}, taking the large-${\cal S}$ 
limit requires that $g$ scales as $\frac{1}{\cal S}$ in order to keep
the corresponding energy finite: such condition turns into $g{\cal S}\sim 
\mathrm{const}$ or, quite equivalently,
\begin{equation}
\label{e.g_appr}
g^{m}{\cal S}^{\ell}\sim 0 \quad \forall m>{\ell} \geq 1~;
\end{equation}
this is how we will hereafter enforce the large-${\cal S}$ condition whenever 
needed.  
We notice that, according with these conditions, the weak-coupling 
constraint, $g\ll1$ and yet finite, introduced in Sec.~\ref{ss.weak-coupling}, corresponds
with taking \sv $\gg 1$ and yet finite.  
Moreover, the above reasoning also applies if the large-${\cal S}$ spin 
operators $\hat S^\alpha$ enter the propagator further multiplied by other
operators acting on the Hilbert space of a different physical subsystem, 
with which they therefore inherently commute, such as the qubits operators 
$\hat\sigma^z_{\A,\B}$ in the second term of Eq.\eqref{e.Htot}.

\subsection{Spin Coherent States} 
\label{ss.SCS} 
Spin coherent states $\ket{\Omega}$ for a system with 
$|\hat{\mathbf S}|^2={\cal S}({\cal S}+1)$,
hereafter indicated by ${\cal S}$CS,  are constructed as (see for 
instance 
Ref.~\cite{ZhangFG90}) 
\begin{equation}\label{e.scs}
\Ket{\Omega}
= e^{\Omega \hat{S}^{-}- \Omega^{*} \hat{S}^{+}} \Ket{0}=\hat{\Omega}\Ket{0}~,
\end{equation}
where $\Omega\in \mathbb{C} $ parametrizes the sphere via
\begin{equation}
\Omega =\frac{\vartheta}{2}e^{i\varphi}~,
\label{e.Omega(theta,phi)}
\end{equation}
with $(\vartheta,\varphi)$ the polar angles,
and $\hat{\Omega}:=\exp\{\Omega \hat{S}^{-}- \Omega^{*} \hat{S}^{+}\}$ 
the so 
called \textit{displacement operator}.
The state $\ket{0}$ is arbitrary, but 
it is most often chosen as one of the eigenstates $\{\ket{M}\}$ 
of $\hat S^z$, typically the one with $M={\cal S}$. This is the choice 
hereafter understood. In Eq.\eqref{e.scs} the state $\ket{0}$ is dubbed 
"reference" state. Notice that the \sv CS depend on the value of $\cal S$ 
(as the acronym suggests) but, 
for the sake of a lighter notation, we will avoid to explicitly write down 
such dependence whenever not misleading.

\sv CS have many properties, some of which are reported in 
Appendix \ref{a.scs}. Particularly relevant to this work
is the one-to-one correspondence between displacement operators and normalized 
vectors in ${\mathbb R}^3$
\begin{equation}
\hat{\Omega}\leftrightarrow\Omega\leftrightarrow
{\mathbf n}(\Omega):=
(\sin\vartheta\cos\varphi,\sin\vartheta\sin\varphi,\cos\vartheta)~,
\label{e.one-to-one_SCS}
\end{equation}
 and 
the composition rule for  displacement operators 
that reads 
\begin{equation}\label{e.composition-displacement}
\hat{\Omega}_1\hat{\Omega}_2 = 
\hat{\Omega}_3e^{i\Phi(\Omega_1,\Omega_2)\hat{S}^{z}},
\end{equation}
where $\Phi(\Omega_1,\Omega_2)$ is a real function, and  
\begin{equation}\label{dyn_versor}
{\mathbf n}(\Omega_3)=\mathbf{R}_{\Omega_1}{\mathbf n}(\Omega_2),
\end{equation}
with
$\mathbf{R}_{\Omega_{1}}$ the rotation in ${\mathbb R}^3$ defined in 
Eq.~\eqref{a.O(3)_SCS}.

The composition rule \eqref{e.composition-displacement} means
that a displacement operator transforms 
any \sv CS into another \sv CS, up to a phase factor.

\section{paramagnetic case}
\label{s.para}

In this section we consider the dynamics of the overall model at $T=0$, 
in the weak-coupling regime, for a 
constant value of the field. Such value is 
understood sufficiently large to guarantee an  
approximately paramagnetic behaviour of $R$ in the absence of 
$D$.

\subsection{Initial state}
\label{ss.initial_state-para}
Consistently with the ring behaving as a paramagnet, we choose its
initial state as
\begin{equation}
\ket{\mathrm{R}}=\otimes_{i=1}^{N}\ket{\uparrow_i}~,
\label{e.ini_state_R-para}
\end{equation}
where $\ket{\uparrow_{i}}$ are the eigenstate of $\hat{s}_{i}^{z}$ with 
eigenvalue $\frac{1}{2}$.
In fact, we will adopt a description in terms of 
spin-$\frac{1}{2}$CS
identifying each $\ket{\uparrow_i}$ with the reference state 
$\ket{0_i}$ used to 
define the spin-$\frac{1}{2}$CS for the particle sitting at site $i$.
For the sake of clarity, these spin-$\frac{1}{2}$ coherent states will 
be hereafter 
indicated 
by 
$\ket{\omega_i}$~.
We therefore write the initial state of the system in the paramagnetic
case as
\begin{equation}
\ket{\Psi_{\rm para}(0)}=\ket{D}
\otimes_{i=1}^{N}\ket{\omega_i=0}~.
\label{e.ini_state-para}
\end{equation}

\subsection{Propagator}
\label{ss.propagator-para}

We handle the propagator via
the Zassenhaus formula \eqref{e.Zassenhaus_left} with
$\lambda=-it$ and $\hat X=\hat{H}_{\rm I}$. This implies that $\hat{Y}$ 
is proportional to $g$, and we can implement the 
weak-coupling constraint 
\eqref{e.weak-coupling} by only taking terms in
Eq.~\eqref{e.Cn+1} which are linear in $g$, thus getting
\begin{equation}
\begin{split}
& C_{n+1} = \frac{g}{2}\frac{(\hat{\sigma}_{A}^{z}+
\hat{\sigma}^{z}_{B})h^{n}}{(n+1)!}
\sum_{i}[(-1)^{n+1}\hat{s}_{i}^{+}-\hat{s}_{i}^{-}]~.
\end{split}
\end{equation}
By carefully manipulating the factors of the Zassenhaus formula,
we get
\begin{equation}
\label{e.propagator-para}
e^{-it\hat{H}}_{\rm para}
\simeq
\left(\prod_{i=1}^{N}e^{\frac{(\hat{\sigma}_{A}^{z}+\hat{\sigma}^{z}_{B})}{2} 
\left[l(t)\hat{s}_{i}^{-} - l(t)^{*}\hat{s}_{i}^{+}\right]} 
\right) 
e^{-it\hat{H}_{I}}~,
\end{equation} 
with
\begin{equation}
l(t)= \frac{g}{h}(1-e^{-ith}).
\label{e.l(t)-para}
\end{equation}

\subsection{Evolved state}
\label{ss.evo_state-para}

The evolved state is obtained by acting with the
propagator \eqref{e.propagator-para} on the initial state 
\eqref{e.ini_state-para}.  
In fact, the form of the above propagator dictates to first evaluate the 
action of $e^{-it\hat{H}_{I}}$ on the initial state of the ring. However, as we 
are in the paramagnetic case, the state (\ref{e.ini_state_R-para}) is a 
good
approximation of the ground state of $\hat{H}_{I}$ with energy 
$E_{\rm gs}$; therefore, the second factor in the r.h.s. of 
Eq.~\eqref{e.propagator-para} gives rise to an irrelevant overall phase 
factor $\exp\{-itE_{\rm gs}\}$ that we will hereafter drop. 
We thus find
\begin{equation}
\label{e.Psi(t)-para}
\ket{\Psi_{\rm para}(t)}
=\sum_{\gamma}c_{\gamma}\ket{\gamma} 
\left(\otimes_{i=1}^{N} 
e^{\pi_{\gamma}\left[l(t)\hat{s}_{i}^{-}-l(t)^{*}\hat{s}_{i}^{+}\right]}
\ket{0_i}\right) ~,
\end{equation}
where $\pi_\gamma$ are the eigenvalues of 
$(\hat\sigma^z_{\A}+\hat\sigma^z_{\B})/2$.
As each exponential in the above expression is the
displacement operator for one spin of the ring acting 
on the respective reference state, it is
\begin{align}\label{e.final_state-para}
&\ket{\Psi_{\rm para}(t)}=\sum_\gamma c_\gamma\
\otimes_{i=1}^N\ket{\omega^\gamma(t)}_i~,\\ 
&{\rm with}~~~~\omega^\gamma(t)=\pi_\gamma l(t)~,
\label{e.omega_gamma(t)}
\end{align}
and $l(t)$ as in Eq.~\eqref{e.l(t)-para}.

\section{diabatic case}
\label{s.diabatic}

In this Section we study the dynamical process underlying our proposal
for protecting internal entanglement by the critical slowing down 
implied by the KZM. We remind that we now consider the model in the 
weak coupling regime, 
with a 
time dependent field $h(t)=h_0-vt$, and $h_0\gsim 1$ 
so as to set $R$ in its disordered critical region, where domains of 
dynamically correlated spins exist according to the phenomenology 
described in Sec.\ref{s.model}. If the ring 
has already entered the diabatic region, $t>\bar t$,
these domains are effectively non-interacting and frozen in size,
their length being on the order of
$\xi_d:=\xi[\epsilon(\bar t)]$, which is the same
as saying that each domain involves  $\xi_d$ 
adjacent spins of the ring, given the 
dimensionless character of all our expressions. Since $R$ is made 
of $N$ spins, the number of distinct domains is
$n_d=N/{\xi_d}$.

\begin{figure}[H]
\centering
\includegraphics[width=0.6\linewidth]{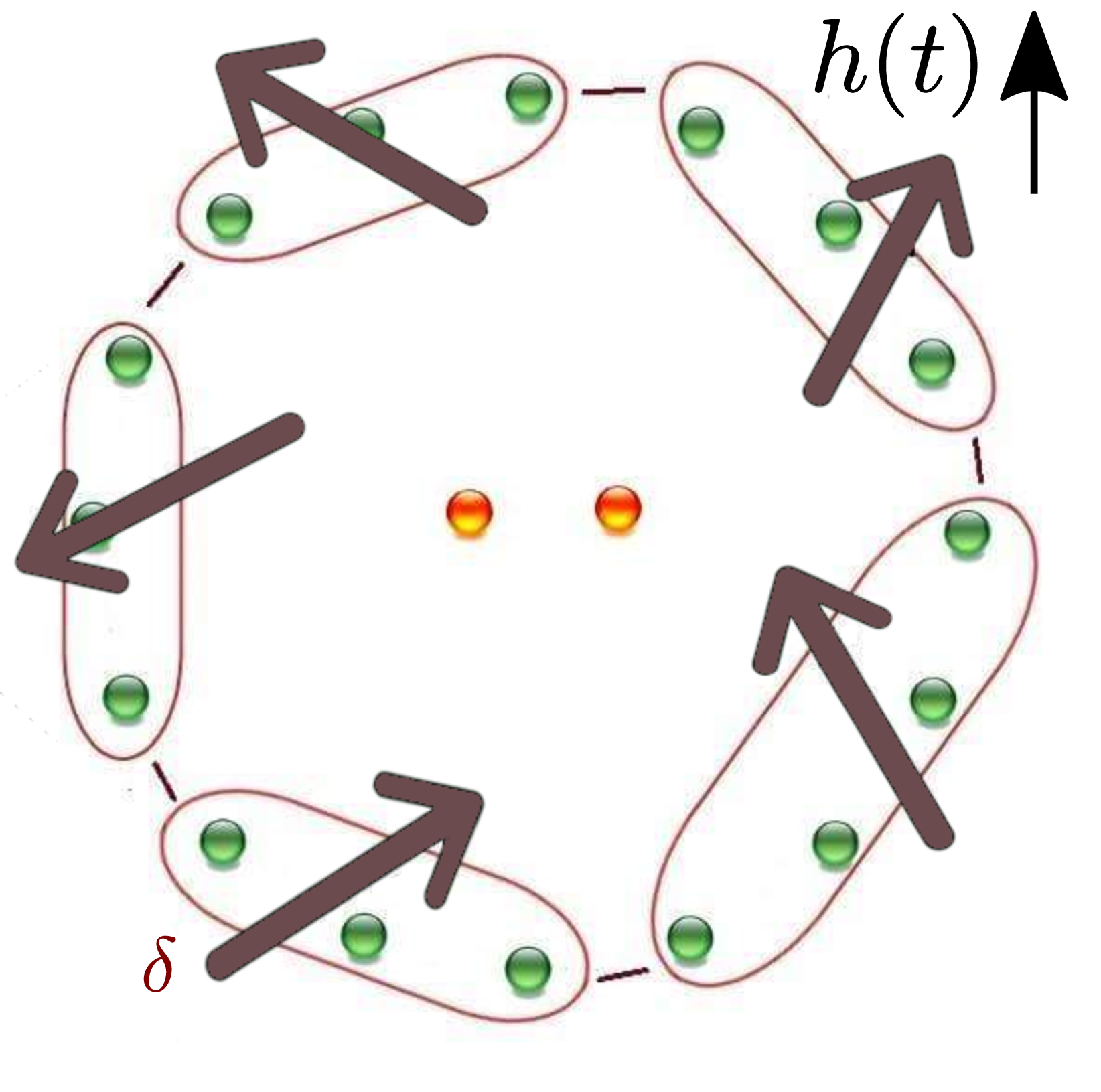}
\caption{\small{Schematic representation of the spins $S_{j}$ with the central qubit pair.}}
\label{fig_domains}
\end{figure}

Spins within the same domain stay roughly aligned with each other by 
definition: therefore, the internal dynamics of each domain can be 
neglected, and the evolution of $R$ can be described in terms of 
spin operators relative to distinct domains. In other terms, 
one can replace the notion of domain as a set of $\xi_d$ 
spin-$\frac{1}{2}$ particles with that of one single spin-${\cal S}_d$ 
system, with ${\cal S}_d\sim\xi_d/2$. Formally, this is done by defining
the collective spin operators  $\hat{\mathbf S}:=
\sum_{i=1}^{\xi_d}\hat{\mathbf s}_{i}$, such that 
$|\hat{\mathbf S}|^2={\cal S}_d({\cal S}_d+1)$, so that the whole ring
 can be described by a set of $n_d$ spin operators
$\{\hat{\mathbf S}_\delta\}$, with $\delta=1,...n_d$, each describing a 
spin-${\cal S}_d$ system, with the same ${\cal S}_d\sim\xi_d/2$, as depicted 
in Fig.\ref{fig_domains}

Once the above description is adopted, the original exchange interaction 
in Eq.~\eqref{e.Hring} is mapped into the effective one 
\begin{equation}
\label{e.domain_Ising} 
-j_{\rm eff}\sum_{\delta=1}^{n_d}\hat{S}_{\delta}^{x}\hat{S}_{\delta+1}^{x}~, 
\end{equation} 
with a dimensionless coupling $j_{\rm eff}$ which is 
determined according to the following reasoning. Given the nearest 
neighbour nature of the original Ising exchange, its contribution as from 
one single domain is $\sim \xi_d/4$, and from the whole ring is $\sim 
n_d\xi_d/4=N/4$, i.e. a constant that can be safely neglected. This 
means that the total exchange energy of the original model must equal 
the interaction energy between neighbouring spins on the edge of adjacent 
domains, i.e $ j_{\rm eff}n_d{\cal S}_d^2 m^2 \simeq n_d 
m^2$,  where we have written $\braket{\hat{s_i}^x\hat s_{i+1}^x}$ as 
$m^2$
by Eq.~\eqref{e.correlations} with $r=1$ and $\xi\gg1$.
In fact, as we are in the 
disordered critical region, it is $m=0$; however, the above reasoning 
works regardless of what side of the QPT is considered, and we can 
safely use it to determine how $j_{\rm eff}$ scales with the domains 
size. Finally,  reminding that ${\cal S}_d \sim\xi_d/2$, we
get 
\begin{equation} j_{\rm eff}\sim\frac{2}{\xi_d^2}\ll 1~.
\label{e.jeff} 
\end{equation} 
The strong reduction of the Ising coupling 
between domains represented by Eq.~\eqref{e.jeff}, is consistent with the 
KZM picture of approximately non interacting domains, and allows us to 
neglect the Ising term in Eq.~\eqref{e.Hring} and write the effective 
Hamiltonian in the diabatic setting as 
\begin{equation}\label{Heff} 
\hat{H}_{\mathrm{dia}}(t)\simeq-h(t)\sum_{\delta=1}^{n_d} 
\hat{S}_{\delta}^{z}- 
\frac{g}{2}(\hat{\sigma}_{{\A}}^{z}+\hat{\sigma}_{{\B}}^{z}) 
\sum_{\delta=1}^{n_d}(\hat{S}_{\delta}^{+}+\hat{S}_{\delta}^{-}). 
\end{equation} 
It is worth noticing that all the operators 
$\{|\hat{\mathbf{S}}_{\delta}|^{2}\}$ commute with $\hat{H}_{\rm 
dia}(t)$ at any time, which formally confirms 
our considering ${\cal S}_{d}$ fixed.

\subsection{Initial state of the ring (diabatic)}
\label{ss.initial-state-dia}

Consistently with the above picture we take the initial state of the ring as

\begin{equation}\label{e.ini_state_R-dia}
\Ket{\mathrm{R}_{\rm dia}}=\otimes_{\delta=1}^{n_d}\Ket{\Delta_{\delta}},
\end{equation}
where $\Ket{\Delta_{\delta}}$ is the initial 
state of 
the 
$\delta$-th domain, that is determined by the following reasoning.
The KZM implies that each domain behaves as a spin-${\cal S}_d$ system,
with ${\cal S}_d\gg1$: given the large value of ${\cal S}_d$ one can 
resort to a semiclassical picture and say that each domain 
points in some direction 
${\mathbf n}_\delta(0):={\mathbf 
n}(\vartheta_\delta(0),\varphi_\delta(0))$. 
On 
the 
other 
hand, there 
exist quantum spin-${\cal S}$ states which are in one-to-one 
correspondence with 
unit vectors in ${\mathbb R}^3$ and that 
formally transform into those 
vectors in the ${\cal S}\to\infty$ limit: they are the ${\cal S}$CS 
introduced in Sec.~\ref{ss.SCS}. Therefore, it makes sense to choose 
\begin{equation}
\label{e.ini_state_R-diaSCS}
\Ket{\Delta_\delta}=\Ket{\Omega_\delta(0)}
= 
e^{\Omega_\delta(0)\hat{S}_\delta^- 
-(\Omega_\delta(0))^*\hat{S}_\delta^+}\Ket{0_\delta},
\end{equation}
where 
$\Omega_\delta(0)$
is in one-to-one correspondence with the above direction 
${\mathbf n}_\delta(0)$ via Eq.~\eqref{e.one-to-one_SCS}.
As for the choice of the set of initial domains directions, i.e. 
of the $n_d$ parameters $\{\Omega_\delta(0)\}$, we have used a 
specific procedure to make it consistent with the expected value of the 
ring magnetization along the $z$ direction, as described in Appendix 
~\ref{a.ini_SCS-dia}.

\subsection{Propagator (diabatic)}
\label{ss.propagator-dia}
The Hamiltonian in the diabatic setting is inherently time-dependent, 
meaning that, at variance with the paramagnetic case considered in 
Sec.~\ref{s.para}, the propagator embodies a troublesome time-ordering 
operator.
However, since
$[H_{\mathrm{dia }}(t_{1}),H_{\mathrm{dia}}(t_{2})] \sim 
v(t_{1}-t_{2})/{\cal S}_d$, and we are dealing with 
extended domains (${\cal S}_d\gg 1$), 
the propagator can still be written as 
$\exp\{-it\hat{H}_{\mathrm{dia}}(t)\}$
as far as $vt$ is not too large, which is guaranteed, via 
Eq.~\eqref{e.epsilon(t)}, by the diabatic 
setting, $h_0\gsim 1$ and $\epsilon(t)>0$.

Therefore, we can again handle the propagator via the Zassenhaus formula 
\eqref{e.Zassenhaus_left} with $\lambda=-it$, now setting
$\hat{X}=-h_t\sum_\delta^{n_d}\hat{S}_{\delta}^{z}$, and 
$\hat{Y}=-\frac{g}{2}(\hat{\sigma}_{\A}^{z}+\hat{\sigma}_{\B}^{z})
\sum_\delta^{n_d}(\hat{S}_{\delta}^{+}+\hat{S}_{\delta}^{-})$,
where $h_t:=h(t)$ for the sake of a lighter notation.
Since ${\cal S}_d\gg 1$ we use the approximation \eqref{e.g_appr} and 
get 
\begin{equation}
\begin{split}
& 
C_{n+1}\sim\frac{(-1)^{n}}{(n+1)!}(it)^{n+1}h_t^{n}
\frac{g}{2}(\hat{\sigma}_{\A}^{z}+\hat{\sigma}_{\B}^{z}) 
\cdot \\ 
& \cdot \sum_\delta[\underbrace{ \hat{S}_\delta^{z},...[\hat{S}_\delta^{z}}_{n-times},\hat{S}_\delta^{+} + \hat{S}_\delta^{-}]...] = \\
& \begin{cases}
& 
\frac{(it)^{n+1}}{(n+1)!}h_t^{n}\frac{g}{2}(\hat{\sigma}_{\A}^{z}
+\hat{\sigma}_{\B}^{z})\sum_\delta(\hat{S}_\delta^{+}+\hat{S}_\delta^{-}) 
\quad \mathrm{if} \, n \, \mathrm{is} \, \mathrm{even} \\
& \quad \\
& \frac{(it)^{n+1}}{(n+1)!}h_t^{n}
\frac{g}{2}(\hat{\sigma}_{\A}^{z} + \hat{\sigma}_{\B}^{z})
\sum_\delta(\hat{S}_\delta^{+} - 
\hat{S}_\delta^{-}) \quad \mathrm{if} \, n \, \mathrm{is} \, \mathrm{odd}.\\
\end{cases}
\end{split}
\end{equation}
By carefully manipulating the factors of the Zassenhaus formula, 
we obtain
\begin{equation}\label{e.propagator_dia}
e_{\mathrm{dia}}^{-it\hat{H}} 
\sim \prod_\delta  e^{ith_t\hat{S}_{\delta}^{z}} 
e^{\frac{(\hat{\sigma}_{\A}^{z}+\hat{\sigma}_{B}^{z})}{2}\left[f(t)\hat{S}_\delta^{-}-f^{*}(t)\hat{S}_\delta^{+} 
\right]} ,
\end{equation}
with
\begin{equation}\label{ft}
f(t)=\frac{g}{h_t}(e^{ith_t}-1)~.
\end{equation}

\subsection{Evolved state (diabatic)}
\label{ss.evolved_state-dia}

Under the effect of the above propagator, the initial state 
\eqref{e.ini_state-separable}, with $\ket{R}$ as from 
Eqs.~\eqref{e.ini_state_R-dia} and \eqref{e.ini_state_R-diaSCS},  
evolves into
\begin{equation}
\label{unex_evstate}
\Ket{\Psi_{\rm dia}(t)} = \sum_{\gamma}c_{\gamma}\Ket{\gamma} 
\otimes_\delta e^{ith_t\hat{S}_\delta^{z}}e^{\pi_{\gamma}
\left[ f(t)\hat{S}_\delta^{-}-f(t)^{*}\hat{S}_\delta^{+}\right]}
\Ket{\Omega_\delta(0)},\\
\end{equation}
where $\pi_{\gamma}$ are the eigenvalues of 
$(\hat{\sigma}_{\A}^{z}+\hat{\sigma}_{\B}^{z})/2$.

To evaluate the action of the propagator on the initial state, we first 
notice that
\begin{equation}\label{interaction_propagator}
e^{\pi_{\gamma}\left[f(t)\hat{S}_{\delta}^{-}-
f(t)^{*}\hat{S}_{\delta}^{+}\right]} 
= 
e^{\Omega^\gamma(t)\hat{S}_{\delta}^{-}
-(\Omega^\gamma(t))^{*}\hat{S}_{\delta}^{+}}
=\hat{\Omega}_\delta^{\gamma}(t)~,
\end{equation}
with 
$\Omega^{\gamma}(t):=\pi_{\gamma}f(t)$, i.e., via 
Eq.\eqref{e.Omega(theta,phi)}, 
\begin{equation}
\begin{split}
\begin{cases}
& 
\vartheta^{\gamma}(t)=
\frac{g|\pi_{\gamma}|2\sqrt{2}}{h_t}\sqrt{(1-\cos(th_t))} 
\\
& \varphi^{\gamma}(t)= \arctan\left[ 
\frac{\pi_{\gamma}g}{h_t}(\cos(th_t)-1),
\frac{\pi_{\gamma}g}{h_t}\sin(th_t) \right].
\end{cases}
\end{split}
\end{equation}
via Eq.~\eqref{e.Omega(theta,phi)}.

Then, using the composition rule \eqref{e.composition-displacement} and 
the definition \eqref{e.ini_state_R-diaSCS}, we obtain
\begin{equation}\label{e.comp_disp-dia}
\hat{\Omega}_\delta^{\gamma}(t)
\ket{\Omega_\delta(0)}=
\ket{\Omega_{\delta}^{\gamma}(t)}e^{i\Phi_\delta^\gamma(t){\cal S}_d}~,
\end{equation}
with 
$\Phi^\gamma_{\delta}(t)=
\Phi(\Omega^\gamma(t),\Omega_\delta(0))\in{\mathbb R}$,
and
\begin{equation}
{\mathbf n}_\delta^\gamma(t)= 
{\mathbf R}_{\Omega^\gamma(t)}{\mathbf n}_\delta(0)~.
\label{e.n_delta_gamma(t)}
\end{equation}

The final state thus reads
\begin{equation}
\label{e.final_state-dia}
\Ket{\Psi_{\rm dia}(t)} = 
\sum_{\gamma}c_{\gamma}\ket{\gamma}
\otimes_{\delta}e^{i\Phi_{\delta}^{\gamma}(t){\cal S}_d}
e^{ith_t\hat S^z_\delta}\ket{\Omega^\gamma_\delta(t)}.
\end{equation}
The further action of the exponential containing $\hat S^z_\delta$ can 
be made explicit via the decomposition \eqref{mz_decomposition} of ${\cal S}$CS on 
the eigenstates of $\hat{S^z}$ reported in Appendix \ref{a.scs}. 
However, as this is irrelevant in what follows, we keep 
the state $\ket{\Psi_{\rm dia}(t)}$ as it is in 
Eq.~\eqref{e.final_state-dia}

\section{Entanglement evolution}
\label{s.entanglement}

In this section we focus upon the internal entanglement featured 
by the evolved states in the 
paramagnetic and diabatic setting, Eqs. \eqref{e.final_state-para} and 
\eqref{e.final_state-dia}, respectively.
We first notice that in both cases it is
\begin{equation} 
\ket{\Psi(t)}=\sum_{\gamma}c_{\gamma}\ket{\gamma}\ket{R^{\gamma}(t)}~,
\end{equation} 
and hence, by partially tracing $\ket{\Psi(t)}\bra{\Psi(t)}$ upon the 
Hilbert space of the ring, the state of the device reads
\begin{equation}
\rho_{D}(t)=
\sum_{\gamma\gamma'}\left(c_{\gamma}c_{\gamma'}^{*}
\ket{\gamma}\bra{\gamma'}\right)
\braket{R^{\gamma'}(t)|R^\gamma(t)}~,
\label{e.rho_D}
\end{equation}
with
\begin{equation}
\braket{R^{\gamma'}(t)|R^\gamma(t)}_{\rm para}=
\prod_{i=1}^{N}\braket{\omega^{\gamma'}_i(t)|\omega^{\gamma}_i(t)}
\label{e.overlap-para}
\end{equation}
in the paramagnetic case, and
\begin{equation}
\braket{R^{\gamma'}(t)|R^\gamma(t)}_{\rm dia}=
\prod_{\delta}^{n_d}
e^{i[\Phi_{\delta}^{\gamma}(t)-\Phi_{\delta}^{\gamma'}(t)]{\cal S}_d}
\braket{\Omega_{\delta}^{\gamma'}(t)|
\Omega_{\delta}^{\gamma}(t)}~
\label{e.overlap-dia}
\end{equation}
in the diabatic one.

To proceed with a quantitative analysis, we must choose a 
specific initial state for the device, and we go for
\begin{equation}
\ket{D}=\frac{1}{\sqrt{2}}\left(\ket{00}+\ket{11}\right)~,
\label{e.Bell}
\end{equation}
which is a maximally entangled state.
This implies that in all our formulas $\gamma$ takes just two values, 
hereafter labelled by $+$ and $-$, corresponding to $\pi_\pm=\pm 1$.
Moreover, we have to evaluate the overlaps between coherent states 
entering Eqs.~\eqref{e.overlap-para} and \eqref{e.overlap-dia}, which we 
do by means of Eq.~\eqref{overlapSCS}.

Finally, a comparison between the time dependence of the internal 
entanglement in the paramagnetic and diabatic settings 
can be developed in terms of the concurrence $C_{AB}(\rho)$
between $A$ and $B$ relative to the state $\rho_D(t)$ in the two cases. 

In the paramegnetic setting we find
\begin{equation}
C_{AB}^{\rm para}(\rho_{D}(t))=\max\left\lbrace 0,  
\cos\left(\frac{\Theta(t)}{2}\right)^{N}\right\rbrace,
\label{e.conc-para}
\end{equation}
with
\begin{align}
\cos(\Theta(t))&:=
\cos\theta^{+}(t)\cos\theta^{-}(t)+\\
+&\sin\theta^{+}(t)\sin\theta^{-}(t)\cos[\phi^{+}(t)-\phi^{-}(t)]~,
\end{align}
with $(\theta^{\pm}(t), \phi^{\pm}(t))$ 
such that $\omega^\pm(t)=\frac{\theta}{2}e^{i\phi}$ and $\omega^\pm(t)$ 
from Eq.~(\ref{e.omega_gamma(t)}).

In the diabatic setting, instead, we get
\begin{equation}
C^{\rm dia}_{AB}(\rho_{D}(t))= 
\max \left\lbrace 0,\left[ 
\prod_{\delta=1}^{n_d}
\cos\left(\frac{\Theta_\delta(t)}{2}\right)\right]^{2{\cal S}_d}\right\rbrace~,
\label{e.conc-dia}
\end{equation}
with
\begin{align}
\cos(\Theta_{\delta}(t))&:=
\cos\vartheta_{\delta}^{+}(t)
\cos\vartheta_{\delta}^{-}(t)+\\
+&
\sin\vartheta_{\delta}^{+}(t)
\sin\vartheta_{\delta}^{-}(t)
\cos[\varphi_{\delta}^{+}(t)-\varphi_{\delta}^{-}(t)]~,
\end{align}
and $(\vartheta_{\delta}^{\pm}(t),\varphi_\delta^\pm)$ from 
Eqs.~\eqref{e.n_delta_gamma(t)} and \eqref{e.one-to-one_SCS}.

\section{Results} \label{s.Results}

Before commenting upon the figures, we recall some important aspects of the model (\ref{Heff}), that concerns the system $R+D$ when the former is in the diabatic critical region. Firstly, we stress that the form of the propagator as in Eq. (\ref{e.propagator_dia}), with
a time-dependent external magnetic field, holds provided that the time interval 
$t$ of evolution from the inital state (\ref{e.ini_state_R-dia}) to the final 
state (\ref{unex_evstate}) satisfies $vt\ll1$. Secondly, the weak-coupling constrain $g\ll 1$ is enforced by the large-${\cal S}$ condition (\ref{e.g_appr}), or quite equivalently the by fact that $g$ scales as $\frac{1}{{\cal S}}$. It is worth saying that we are interested in keeping the interaction between $R$ and $D$ finite, this implying that we will consider spin domain of size $\mathcal{S}\gg 1$ but still finite in order to let the two systems interact. Finally, in the following we take $\xi_{0}=1$. 

\begin{figure}
\includegraphics[width=\linewidth]{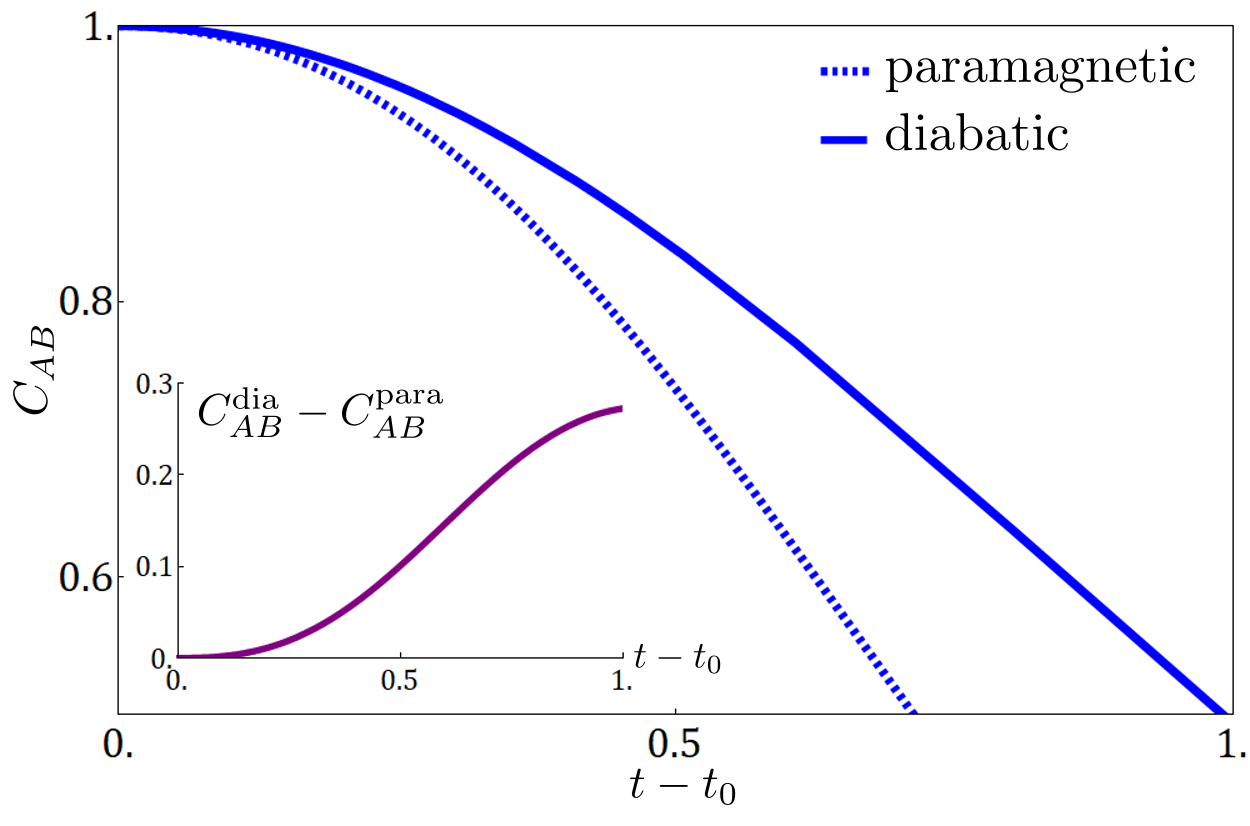}
\caption{\small{The blue solid line represents $C_{AB}(t)$ in the diabatic region with $h_{0}=1.01$, the blue dashed line represents $C_{AB}(t)$ in the paramagnetic region with $h=2$. We set $N=120$, $v= 0.6 \cdot 10^{-3}$ and $t_0=\bar t+12$. The inset shows the difference between the two lines.}}
\label{fig:f_dia-vs-para}
\end{figure}

In Fig. \ref{fig:f_dia-vs-para} we show $C_{\A\B}$ as a function of $t\in[0,1]$, 
for a ring of $N=120$ sites.
In the paramagnetic setting we take $h=2$, while in the diabatic one 
we choose values of the parameters consistent with the simplest 
non-trivial situation, i.e. $n_d=2$, $h_0=1.01$, $v=6\cdot 10^{-4}$ and 
$t_0=\bar t+12$.

We see that the entanglement of the qubit pair assumes its 
maximum value at the initial time, consistently with the choiche of the 
initial state (\ref{e.Bell}) and then decreases, due to the 
interaction between the qubit pair and the ring. However, the decline of 
the concurrence is slower in the diabatic setting than in the 
paramagnetic one, as displayed in the inset of the figure that shows the difference between the concurrence in the diabatic region and the one in the paramgnetic region.  

We then focus our attention on the evolution of the concurrence inside 
the diabatic region for different choices of $v$ and $N$
fixed.
This means that we are comparing the same 
microscopic model, corresponding to the Hamiltonian (\ref{Heff}), for 
systems that differ in the number $n_d$ and the size $\xi_d$ of 
the domains.
Specifically, as the speed $v$ decreases, the 
ring splits into a decreasing number of larger and larger domains. 
Furthermore, we prepare the ring so that the evolution we are interested 
in starts inside the diabatic region, at the initial time $t_{0} \geq 
\bar{t}$. The boundary $ \bar{t}$ of such region depends on the speed 
$v$, as shown in Eq. (\ref{e.bart}), and thus also the value of the 
magnetic field at the time $\bar{t}$, $h(\bar{t})$.

\begin{figure}
\includegraphics[width=\linewidth]{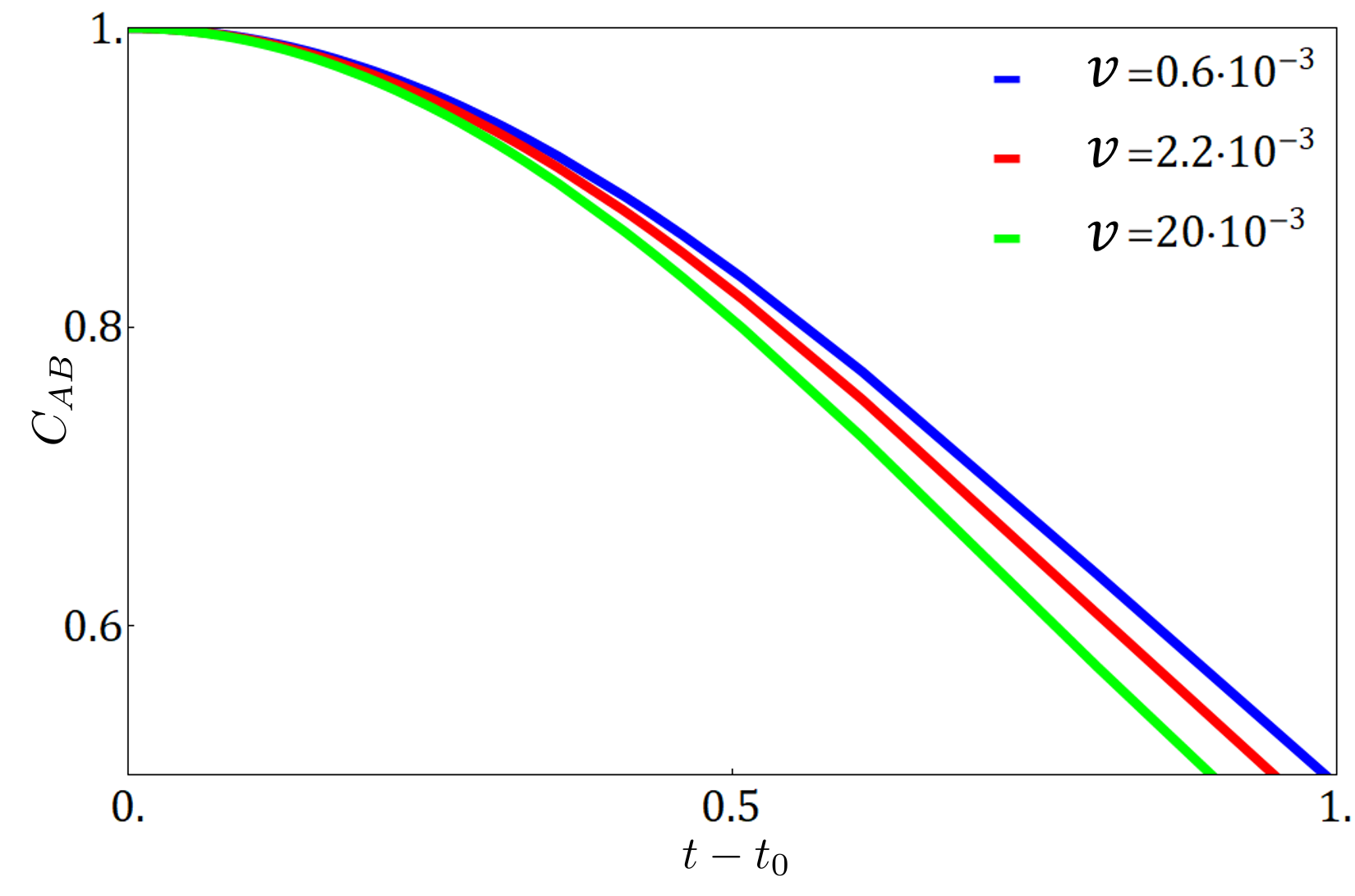}
\caption{\small{Concurrence $C_{AB}(t)$ in the diabatic region vs time interval $t-t_0$, with $N=120$ and $g=1/6$. The values of the parameters are $v=0.6\cdot 10^{-3}$, $t_0=\bar t+12$, 
$ \xi_{d}=60$ and $h_{0}=1.01$ for the blue line, $v=2.2\cdot 10^{-3}$, $t_0=\bar t+1.5$, 
$ \xi_{d}=30$ and $h_{0}=1.03$ for the red line,  $v=20\cdot 10^{-3}$, $t_0=\bar t+0.5$, 
$ \xi_{d}=10$ and $h_{0}=1.09$ for the green 
line.}}
 \label{fig:diabatic}
\end{figure}

%\begin{figure}
%\includegraphics[scale=0.35]{figs/fig2a.pdf}(a)
%\includegraphics[scale=0.2]{figs/fig2a_legend.pdf}
%\includegraphics[scale=0.3]{figs/fig2b.pdf}(b)
%\includegraphics[scale=0.2]{figs/fig2b_legend.pdf}
%\caption{\small{(a) Concurrence $C_{Q_{1}Q_{2}}$ in function of $t$ and $g$, with $N=120$, $\tau_{Q}^{-1}=0.6\cdot 10^{-3}$, $\hat{\xi}=60$ and $h_{0}=1.01$. (b)Contour-plot of $C_{Q_{1}Q_{2}}$. Different colours refer to differents value of $C_{Q_{1}Q_{2}}$, decreasing from white $(C_{Q_{1}Q_{2}} \sim 1)$ to blue $(C_{Q_{1}Q_{2}} \sim 0)$ }}
% \label{fig:fig3}
%\end{figure}
%
%\begin{figure}
%\includegraphics[scale=0.35]{figs/fig3a.pdf}(a)
%\includegraphics[scale=0.2]{figs/fig2a_legend.pdf}
%\includegraphics[scale=0.35]{figs/fig3b.pdf}(b)
%\includegraphics[scale=0.2]{figs/fig3b_label.pdf}
%\caption{\small{(a) Concurrence $C_{Q_{1}Q_{2}}$ in function of $t$ and $g$, with $N=120$, $\tau_{Q}^{-1}=0.6\cdot 10^{-3}$, $\hat{\xi}=60$ and $h_{0}=1.01$. The red lines shows $C_{Q_{1}Q_{2}}$ as a functions of time, with $g= \lbrace 0.1,1,0.9 \rbrace$ (b)Contour-plot of $C_{Q_{1}Q_{2}}$. Different colours refer to different values of $C_{Q_{1}Q_{2}}$, decreasing from white $(C_{Q_{1}Q_{2}} \sim 1)$ to blue $(C_{Q_{1}Q_{2}} \sim 0)$ }}
% \label{fig:fig4}
%\end{figure}

The data displayed in Fig.~\ref{fig:diabatic} were obtained taking $N=120$ and setting 
the value of $g$ equal to $\frac{1}{6}$, which assures the weak-coupling condition 
holds: $C_{AB}$ is reported for different values of the speed  
$v=\lbrace 0.6\cdot10^{-3}, 2.2\cdot 10^{-3}, 20 \cdot 
10^{-3} \rbrace$, which correspond to $R$ being described by spins 
$\mathcal{S}$ of size $\xi_d=\lbrace 60,30,10 \rbrace $ and number
$n_{d}=\lbrace 2,4,12\rbrace$, respectively; as for the initial values 
of the magnetic field we take $h_{0}=\lbrace 1.01, 1.03, 1.09\rbrace$. 
We see that as the environment approaches more and more the critical 
point at the starting time of the dynamics $t_0$, and the number of domains 
decreases, while they grow in size, it behaves more and more macroscopically, 
and the time-decay of entanglement shared between the two qubits slows down 
accordingly.

\begin{figure}
\includegraphics[width=\linewidth]{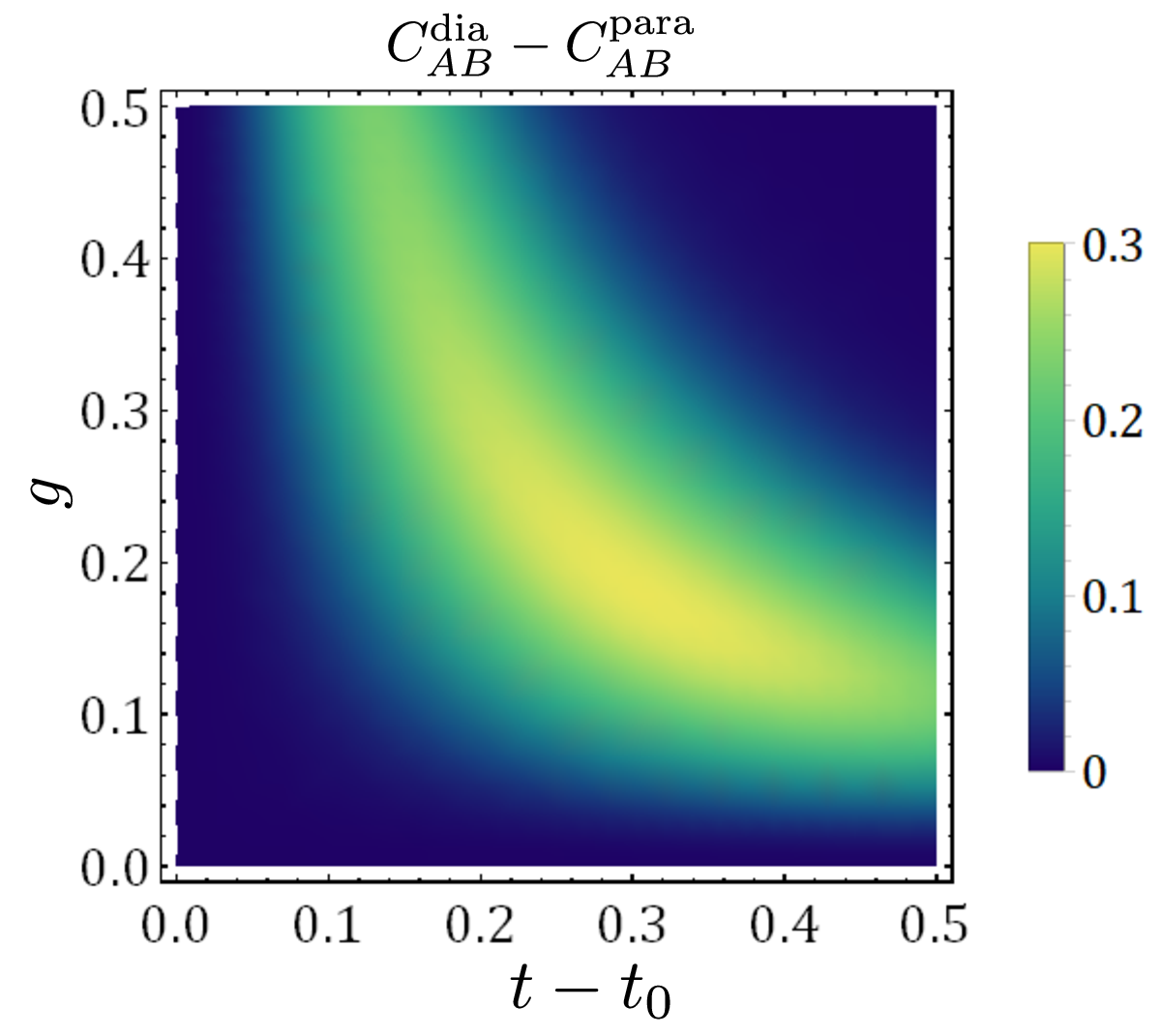}
\caption{\small{Difference between the concurrence $C_{AB}$ in the diabatic and 
the paramagnetic case vs $g$ and $t-t_0$ for $N=1000$;
$h=5$ in the paramagnetic case, while data for the diabatic dynamics 
are obtained for $v=5\cdot 10^{-5}$, $t_0=\bar t$, $ \xi_{d}=200$ 
and $h_{0}=1.001$. }}
\label{fig:diabatic_vs_adiab_density}
\end{figure}

The advantage of working in the critical region is better 
appreciated in Fig. \ref{fig:diabatic_vs_adiab_density}, where the difference 
between the concurrence in the diabatic and the paramagnetic 
regime is shown as a function of coupling and time: a sensible 
entanglement protection is observed for an extended time interval when 
$g$ is on the order of $0.1\div0.2$.  
  
\section{Conclusions}
\label{s.Conclusions}

Our analysis shows that
the critical slowing down observed in the proximity of a QPT
of a many-body system is an effective tool for protecting the
entanglement between components of a quantum device 
when the many-body system acts as the surrounding environment 
of the device. 
In particular, we have considered an environment modelled by an Ising 
chain in transverse field in order to relate this work with the 
availability of experimental evidence that such a model is amenable of 
physical realization 
\cite{ColdeaTWWPTHSK10,SuzukiInoueChakrabart13_book,LiangKKM-QCO15}: 
Indeed, signatures of quantum critical behaviour are seen to persists 
at finite temperature \cite{ColdeaTWWPTHSK10,LiangKKM-QCO15,BayatEtal16,CampbellEtal17,FogartyEtal17}, and can be 
recognized in the behaviour of rather small Ising rings
\cite{OrieuxBBPM14,XueZB17}, meaning that the idea of exploiting 
critical features of the environment in designing apparatuses 
that embody quantum devices might be experimentally tested.

\begin{acknowledgments}
PV gratefully acknowledges support from the Simons Center for Geometry and Physics, 
Stony Brook University, at which some of the research for this paper was 
performed. Financial support from the University of Florence in the 
framework of the University Strategic Project Program 2015 (project 
BRS00215) is gratefully acknowledged. This work is done in the framework 
of the Convenzione operativa between the Institute for Complex Systems 
of the Consiglio Nazionale delle Ricerche (Italy) and the Physics and 
Astronomy Department of the University of Florence.
\end{acknowledgments}

\bibliography{elifi} 

\appendix
\section{Spin coherent states} 
\label{a.scs} 
Spin coherent states can be introduced by following the steps given in 
Ref.~\onlinecite{ZhangFG90}. The first step is the recognition of the 
dynamical group pertaining to the spin system at hand: Since the 
Hamiltonians in Eqs. (\ref{e.propagator-para}) and (\ref{Heff}) are 
linear functions of the operators $\lbrace \hat{s}_{i}^{z}, 
\hat{s}_{i}^{\pm} \rbrace $ and $\lbrace \hat{S}_{\delta}^{z}, 
\hat{S}_{\delta}^{\pm} \rbrace $, respectively, the group 
\cite{Gilmore72, Perelomov72} is $G=SU(2)$. The Hilbert space 
$\mathcal{H}_{S}$ associated to a spin ${\cal S}$, whose Hamiltonian is 
a linear combination of the $SU(2)$ generators, i.e. $\lbrace 
\hat{S}^{z}, \hat{S}^{\pm} \rbrace $, is spanned by $\lbrace 
\Ket{\mathcal{S} ,M} \rbrace $, where $\Ket{\mathcal{S} ,M}$ are 
simultaneous eigenstates of $\hat{\mathbf{S}}^{2}$ and $\hat{S}^{z}$ (in 
order to lighten the notation, we omit the index "$i$" or "$\delta$" 
throughout this appendix, as it does not affect the construction of the 
\sv CS). The reference state \cite{ZhangFG90} is usually taken to be the 
highest- or lowest-weight state of $SU(2)$; the most natural choice is 
the former, i.e. $\Ket{\mathcal{S},\mathcal{S}} \equiv \Ket{0}$. The 
reference state identifies the maximal stability subgroup $H=U(1) $, 
whose elements $\hat{h}$ leave $\Ket{\mathcal{S},\mathcal{S}}$ invariant 
up to a phase factor, according to the form $ \hat{h} =e^{i \alpha 
\hat{S}^{z}} $, $ \alpha \in \mathbb{R}$. The quotient group is thus 
$G/H=SU(2)/U(1)$, which is associated with the two-dimensional sphere 
${\rm S}^{2}$. The \sv CS are eventually defined as

\begin{equation}\label{SU(2)cS}
\Ket{\Omega}=\hat{\Omega} \Ket{0} = e^{\Omega \hat{S}^{-}- \Omega^{*} \hat{S}^{+}} \Ket{0},
\end{equation}
where $\hat{\Omega}$ is referred to as the \textit{displacement operator}; $\Omega \in \mathbb{C} $ parametrizes the sphere ${\cal S}^{2}$ and can be written as a function of the more familiar polar angles $(\vartheta, \varphi)$ as $ \Omega =\frac{\vartheta}{2}e^{i \varphi}$. We notice that Eq.~(\ref{SU(2)cS}), with the definition of the parameter $\Omega$, establishes a one-to-one correspondence between the \sv CS, the elements of the quotient space $G/H$, and the points on the sphere. 

With the previous parametrization, the $SU(2)$ representation of any element $\hat{\Omega}  \in SU(2)/U(1)$ is

\begin{equation}
\mathbf{\Omega}(\vartheta, \varphi)=\begin{pmatrix}
\mathrm{cos}\frac{\vartheta}{2} & -\mathrm{sin}\frac{\vartheta}{2}e^{-i\varphi}\\
\mathrm{sin}\frac{\vartheta}{2}e^{i\varphi} & \mathrm{cos}\frac{\vartheta}{2}
\end{pmatrix}.
\end{equation}   
As shown in Ref.~\cite{CombescureD12}, from the relation between the groups $SO(3)$ and $SU(2)$, it is possible to obtain the representation of $\hat{\Omega}$ in $SO(3)$, which is
\begin{equation}\label{a.O(3)_SCS}
\begin{split}
& \mathbf{R}_{\Omega(\vartheta, \varphi)} = \\
& \begin{pmatrix}
\mathrm{cos}\vartheta\mathrm{cos}^{2}\varphi+\mathrm{sin}^{2}\varphi & \mathrm{sin}\varphi\mathrm{cos}\varphi(1-\mathrm{cos}\vartheta) & -\mathrm{sin}\vartheta\mathrm{cos}\varphi \\
\mathrm{sin}\varphi\mathrm{cos}\varphi(1-\mathrm{cos}\vartheta) & \mathrm{cos}\vartheta\mathrm{sin}^{2}\varphi + \mathrm{cos}^{2}\varphi & \mathrm{sin}\vartheta\mathrm{sin}\varphi \\
\mathrm{sin}\vartheta\mathrm{cos}\varphi & -\mathrm{sin}\vartheta\mathrm{sin}\varphi & \mathrm{cos}\vartheta \\
\end{pmatrix}.\\
\end{split}
\end{equation}
We remind some properties of the \sv CS which turn to be useful in our calculation, taking a specific dimension $2\mathcal{S}+1$ of the Hilbert space.  
First of all, \sv CS are in general not orthogonal, in fact it is
\begin{equation}\label{overlapSCS}
\begin{split}
&|\braket{\Omega^\prime|\Omega}|^{2}= \left( \frac{1+\mathbf{n}(\Omega^\prime) \cdot \mathbf{n}(\Omega)}{2} \right)^{2\mathcal{S}}=\mathrm{cos}^{4\mathcal{S}}\frac{\Theta}{2},\\
\end{split}
\end{equation}
where $\mathbf{n}(\Omega)=(\mathrm{sin}\vartheta\mathrm{cos}\varphi, 
\mathrm{sin}\vartheta\mathrm{sin}\varphi, \mathrm{cos}\vartheta)$ is the 
unit vector along the direction defined by the spherical, polar angles 
$(\vartheta,\varphi)$, while 
$\Theta=\mathrm{cos}\vartheta\mathrm{cos}\vartheta'+\mathrm{sin}\vartheta\mathrm{sin}\vartheta'\mathrm{cos}(\varphi-\varphi')$. 
Nevertheless, the normalization of \sv CS is guaranteed 
$\braket{\Omega|\Omega}=\braket{0|\hat{\Omega}^{\dagger}\hat{\Omega}|0}=\braket{0|0}=1$, 
and the \sv CS become almost orthogonal for large ${\cal S}$, as 
$\lim_{S\to\infty} 
|\braket{\Omega'|\Omega}|^{2}\propto\delta(\Omega-\Omega')$. The 
resolution of the identity reads
\begin{equation}
\hat{\mathbb{I}}= \int d\mu(\Omega)\Ket{\Omega}\Bra{\Omega}= \frac{2\mathcal{S}+1}{4\pi}\int_{\mathcal{S}^{2}} d\Omega\Ket{\Omega}\Bra{\Omega},
\end{equation}
where $d\Omega$ is the solid-angle volume element on ${\rm S}^{2}$, 
namely $d\Omega=\mathrm{sin}\vartheta d\vartheta d\varphi$. Any \sv CS can 
be expanded on the basis $\lbrace \Ket{\mathcal{S},M} \rbrace$
\begin{equation}\label{mz_decomposition}
\Ket{\Omega}=\sum_{M=-\mathcal{S}}^{+\mathcal{S}}g_{M}(\Omega)\Ket{\mathcal{S},M},
\end{equation}
where $g_{M}(\Omega)=\Braket{\mathcal{S},M| \Omega}$ and 

\begin{equation}
g_{M}(\Omega)=\sqrt{\begin{pmatrix} 2\mathcal{S} \\ \mathcal{S}+M \end{pmatrix}} \left( \mathrm{cos}\frac{\vartheta}{2} \right)^{\mathcal{S}+M}\left( \mathrm{sin}\frac{\vartheta}{2} \right)^{\mathcal{S}-M}e^{i(\mathcal{S}-M)\varphi}
\end{equation}
holds.

Finally, the composition-law for different displacement operators is 
needed. To this aim, let us consider the operators $\hat{\Omega}_{1}$ 
and $\hat{\Omega}_{2}$, which are associated to the unit vectors on 
the sphere $\mathbf{n}(\Omega_1)$ and $\mathbf{n}(\Omega_2)$, respectively. It is

\begin{equation}\label{displacement}
\hat{\Omega}_{1}\hat{\Omega}_{2} = \hat{\Omega}_{3}e^{-i\Phi(\Omega_{1},\Omega_{2})\hat{S}^{z}},
\end{equation}
where $\hat{\Omega}_{3}$ is associated to the unit vector 
$\mathbf{n}(\Omega_3)$, obtained from $\mathbf{n}(\Omega_2)$ after the rotation 
$\mathbf{R}_{\Omega_1} \in SO(3)$ induced by the operator 
$\hat{\Omega}_1$, i.e.

\begin{equation}\label{dyn_versor}
\mathbf{n}(\Omega_3)=\mathbf{R}_{\Omega_1}\mathbf{n}(\Omega_2),
\end{equation}
meaning that a displacement operator $\hat{\Omega}$ transforms any \sv CS $\Ket{\Omega'}$ into another \sv CS, up to a phase factor. 

\section{Initial state of domains in the diabatic case}
\label{a.ini_SCS-dia}

The initial state of the ring in the diabatic region in Eqs.~(\ref{e.ini_state_R-dia}) and (\ref{e.ini_state_R-diaSCS})
describes each domain pointing in some direction $ {\mathbf n}_\delta(0):={\mathbf 
n}(\vartheta_\delta(0),\varphi_\delta(0))$, where the spherical polar angles 
($\vartheta_\delta(0), \varphi_\delta(0)$) identify a point on a sphere. 

Referring to the strategy for choosing the initial conditions, it is worth saying that the lack of 
correlations among domains will allow for an independent choice of 
$(\vartheta_\delta(0),\varphi_\delta(0))$. 

Nevertheless their values has to be 
related to the phenomenology of the KZM in the diabatic region. Indeed, we must properly take into account that the magnetization of each domain is proportional to the expectation value of the operator 
$\hat{\mathbf{S}}_{\delta}$ on the state of the $\delta$-th domain, averaged on 
different possible configurations, $\mathbf{M}_{\delta}(t)= 
\braket{\hat{\mathbf{S}}_{\delta}}$, where the time-dependence of 
magnetization is due to the time-evolution of the state of each domain.

The choice of $\vartheta_{\delta}(0)$ is related to the value $h_{0}\equiv 
h(t_0)$ of the external magnetic field at the time $t_0$ (the time when the two-qubit system starts to evolve after having 
been prepared in a well defined initial state), within the 
diabatic region: In fact, since 
$\vartheta_{\delta}(0)$ represents the angle between the unit vector 
${\mathbf n}_\delta$ defined by the pair ($\vartheta_\delta,\varphi_\delta$) and the $z$ 
axes, then $\vartheta_{\delta}(0)$ determines the magnetization of the $\delta$-th 
domain along the $z$ direction. 

We thus select each $\vartheta_{\delta}(0)$ 
in such a way that the average magnetization of the entire ring is equal 
to $n\,m_0^z$, where $m_0^z$ is the equilibrium average magnetization per 
particle of the ring for the chosen value $h_0$ of the external field, as given by 
Ref.~\cite{Pfeuty70}. 

\begin{figure}[H]
\centering
\includegraphics[scale=0.15]{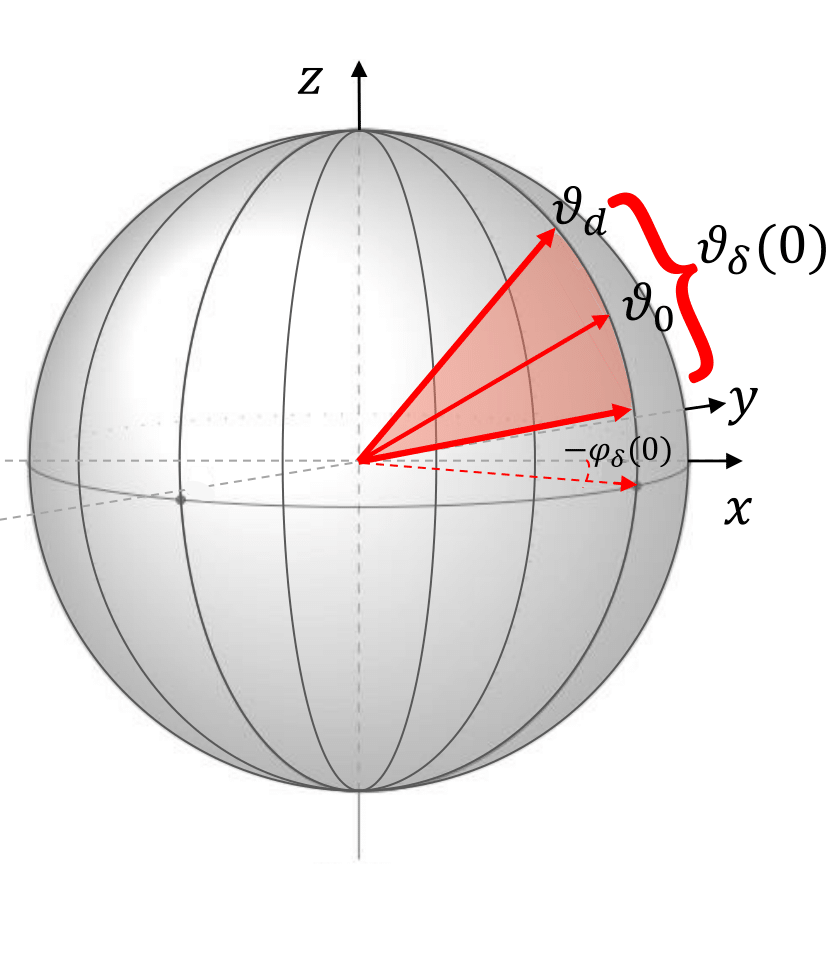}
\caption{Schematic representation of the selection process for the initial point $(\vartheta_\delta(0),\varphi_\delta(0))$.}
\label{f:initial_condition-dia}
\end{figure}

The selection proceeds as follows: we start 
by choosing a domain $\delta_1$, and we select the corresponding 
$\cos(\vartheta_{\delta_1}(0))\equiv m_{\delta_1}^z$ from a uniform distribution 
centered on $m_0^z$ and having width $\Delta_0=|m_d^z-m_0^z|$, $m_d^z$ 
being the magnetization corresponding to the value $h_d=h(\bar t)$ of the external 
field when the ring enters the diabatic region; we then move to another 
domain, $\delta_2$, and we select the corresponding 
$\cos(\vartheta_{\delta_2}(0))$ from a uniform distribution centered on 
$m_1^z=(n\,m_0^z-m_{\delta_1}^z)/(n-1)$ and having width $\Delta_1={\rm min} 
(|m_d^z-m_1^z|,|m_0^z-\Delta_0-m_1^z|)$, and so on up to the last 
$\vartheta_{\delta_n}(0)$. The selection process of $\vartheta_{\delta_1}(0)$ is sketched in Fig. \ref{f:initial_condition-dia}, where $\vartheta_0, ~\vartheta_d$ are defined such that $\cos(\vartheta_0)= m_0$, $\cos(\vartheta_d)= m_d$.

It is worth noting that the choice of such initial 
condition allows us to distinguish different time instants in the 
diabatic region, due to the dependence of the $z$ magnetization on the 
external magnetic field.

As for $\varphi_{\delta}(0)$, i.e. the angle between the projection of the 
unit vector ${\mathbf n}_\delta$ on the $x-y$ plane and the $x$ axes, we have to 
refer to the magnetization along the $x$ direction, which is the order 
parameter of the model. Since we are preparing the ring in the 
disordered region, we choose each $\varphi_{\delta}(0)$ to assume randomly 
the value $0$ or $\pi$, so that the $x$ magnetization of each domain 
can be aligned along the direction $+x$ or $-x$ with the same 
probability.

\end{document}